# Ultrafast measurement of field-particle energy transfer during chorus emissions in space

C. M. Liu[1,2], B. N. Zhao[1,2], J. B. Cao[1,2,†], C. J. Pollock[3], C. T. Russell[4], Y. Y. Liu[1,2], X. N. Xing[1,2], P. A. Linqvist[5], and J. L. Burch[6]

[1]School of Space and Environment, Beihang University, Beijing, China

[2]Key Laboratory of Space Environment Monitoring and Information Processing, Ministry of Industry and Information Technology, Beijing, China

[3]Denali Scientific, San Antonio, TX, USA

[4]Department of Earth and Space Sciences, University of California, Los Angeles, CA, USA

[5]KTH Royal Institute of Technology, Stockholm, Sweden

[6]Southwest Research Institute, San Antonio, TX, USA

# Abstract

Chorus is one of the strongest electromagnetic emissions naturally occurring in space, and can cause hazardous radiations to humans and satellites[1-3]. Although chorus has attracted extreme interest and been intensively studied for decades[4-7], its generation and evolution remain highly debated, due to the complexity of the underlying physics and the limited capacity of previous spacecraft missions[7]. Chorus has also been believed to be governed by planetary magnetic dipolar fields[5,7]. Contrary to such conventional expectation, here we report unexpected observations of chorus in the terrestrial neutral sheet where magnetic dipolar effect is absent. Using unprecedentedly high-cadence data from NASA's MMS mission, we present the first, ultrafast measurements of the wave dispersion relation and electron three-dimensional distributions within the waves, showing smoking-gun evidences for chorus-electron interactions and development of electron holes in the wave phase space. We estimate field-particle energy transfer inside the waves and find that the waves were extracting energy from local thermal electrons, in line with the waves' positive growth rate derived from instability analysis. Our observations, opening new pathways for resolving long-standing controversies regarding the chorus emissions, are crucial for understanding nonlinear energy transport ubiquitously observed in space and astrophysical environments.

Naturally-occurring electromagnetic waves are ubiquitous in the universe[8], which is filled charged particles (e.g., protons and electrons) constituting plasmas, the so-called fourth state (in addition to solid, fluid, and gas) of matter in nature. Since the collision lengths of the charged particles in space are much longer than their free moving lengths, the charged particles are typically collisionless, without direct energy exchange among different particles via binary collisions. The interactions between electromagnetic waves and charged particles, therefore, have been thought to be the most important way that establishes an energy transfer chain among different particle populations, and are usually invoked to explain particle acceleration and scattering observed in laboratory, space, and astrophysical plasmas[1-3,9,10].

Among these naturally-occurring waves, whistler-mode chorus waves—named by their audio-converted sounds resembling chirping chorus of birds at dawn—have attracted extreme interest and been intensively studied for over 70 years[4-7]. Chorus, widely observed in geospace and other planets (e.g., Mars, Jupiter, and Saturn)[11,12], have been documented to play a crucial role in accelerating relativistic electrons in radiation belts[13-15], scattering electrons into the atmosphere to form diffuse and pulsating auroras[16-18], generating hiss waves which can reform the radiation belt[19,20], and mediating the magnetosphere-ionosphere coupling[21]. The most remarkable feature of chorus revealed by satellite observations is their discrete, chirping elements whose central frequency rapidly changes over time, with frequency rising called rising-tone and otherwise falling-tone[8,9]. Their generation, propagation, and associated wave-particle interaction have hitherto been extensively investigated in the context of planets' magnetic dipolar field, with various approaches including numerical simulations and satellite observations[22-33].

Despite the decades-long research, many important issues regarding chorus waves, such as their generation and chirping element formation, remain highly debated. One main reason is that direct measurement of wave-particle energy transfer inside the chorus waves has been lacking, because: 1) chorus waves have been traditionally believed to be closely related to magnetic dipolar field of the planets and have been targeted only in near-planet space, at typical $L$-shells less than 10 $R_E$ (Earth radii)[23, 30]; 2) previous satellites in the near-planet region cannot fully resolve the three-dimensional velocity distributions of resonant electrons in sufficiently-high cadence, due to the presence of strong magnetic field and high-energy electrons therein.

In this study, we fill this gap by presenting the first observation of repetitive, rising-tone chorus waves in the terrestrial midtail neutral sheet ($L = 26\ R_E$), where the magnetic field is dominated by strongly-stretched magnetic field lines rather than magnetic dipolar field (Figure 1). This observation was made by NASA' Magnetospheric Multiscale (MMS) satellites, which were launched in 2015 and equipped with state-of-the-art instruments[35]. In particular, the fast plasma investigation (FPI) instrument of the MMS satellites can measure three-dimensional electron distribution functions orders of magnitude faster than previous instruments, with burst-mode temporal resolution of 30 ms[36]. MMS mission was designed to study electron-scale physical processes, primarily magnetic reconnection in the magnetosphere, which are also relevant to the electron-scale chorus waves. Using MMS' unprecedentedly high-cadence data, we fully resolve the wave dispersion relation and present direct measurement of field-particle energy exchange inside the chorus, showing that electrons were losing energies to the waves. Our observations open new pathways for studying the generation and evolution of the chorus waves, and pose new observational constraints for numerical and theoretical studies of nonlinear wave-particle interactions in collisionless plasmas.

## Results

**Event overview**

On 10 August 2019, the four MMS spacecraft, in a tetrahedral formation separated by ~44 km, were located in the terrestrial mid tail, at [-19.1, 8.5, 4.7] $R_E$ (Earth radii, 6370 km) in the Geocentric Solar Magnetospheric (GSM) coordinates. The time interval discussed for the chorus waves event extends from 15:00:31 to 15:00:36 Universal Time (UT), during which the spacecraft were cruising inside the plasma sheet where hot plasma populations dominate (ion temperature $T_i \approx 4$ keV, electron temperature $T_e \approx 2$ keV, Figs. 2a, 2b, 2g and 2h). During this interval, MMS observed an earthward plasma jet ($V_{i,x} \approx 300$ km s$^{-1}$, Fig. 2D). Inside the jet, a magnetic dip (MD) structure, featuring a weak decrease of local magnetic field strength (from ~4.7 to ~3.3 nT, Fig. 2c), was observed from 15:00:33.2 to 15:00:34.2 UT. The magnetic field $B_x$ component within the MD is basically steady and close to 1.6 nT, indicating that MMS was located inside the neutral sheet. During the MD interval, the magnetic field $B_z$ component dereases (from ~3 to ~0 nT, Fig. 2c), while the magnetic field $B_y$ component increases (from ~3 to ~5 nT, Fig. 2c), indicative of local magnetic field rotation.

The MD is associated with two noticeable changes in particles and fields: 1) during the MD interval, electron temperature anisotropy increases, with electron perpendicular temperature dominating; 2) clear fluctuations in both magnetic and electric fields are seen, suggesting the presence of electromagnetic waves therein. The wave signal is more prominent in the wave spectrograms, where intense wave emissions with frequency extending from near 0.1 $f_{ce}$ ($f_{ce}$ is local electron gyrofrequency) to 0.5 $f_{ce}$, are observed (Figs. 2j and 2k). More interestingly, the waves host discrete, risng-tone elements with chirping rate of ~250 Hz/s, reminiscent of chorus waves observed in geospace[23,30]. Considering the locally enhanced electron perpendicular temperature anisotropy, the waves could be locally-generated chorus waves.

**Wave properties**

To disgnose the wave properties, we first analyze the wave polarizations via the widely-used singular value decomposition (SVD) method. As shown in Figure 3, the waves have large positive ellipticities (~1, Fig. 3c), small propagation angles with respect to magnetic field (< 20°, Fig. 3d), as well as large planarity (~1, Fig. 3e), indicating that they are right-handed polarized, parallel-propagating whistler-mode waves. Considering the presence of the frequency-chirping elements, the electromagnetic emissions are indeed chorus waves.

Taking advantage of the MMS high-cadence data, we calculate the wave vector **k** by utilizing the Bellan method which is based on Ampere's law *(37)*, as shown in Fig. 3g. The obtained **k** correponding to the maximum wave power is 2.2·10$^{-5}$×[-0.31, -0.67, -0.67] m$^{-1}$, roughly antiparallel to local magnetic field, consistent with the small normal angles derived from the SVD analysis. The corresponding parallel wave length is ~280 km, or equivalently 7 $\rho_e$, where $\rho_e$ is local electron gyroradius. The wave phase speed is 3939 km s$^{-1}$, close to 0.22 $V_{te}$ (electron thermal speed). We have also estimated the wave speed using four-spacecraft timing method, which yielded similar results (not shown).

The wave dispersion relation (DR) of the chorus waves resolved from Ampere's law is compared with the theoretical DR for whistler waves in cold plasma (black dashed line in Fig. 4g). The observed DR of the chorus waves is well consistent with the DR of the whistler waves. We further perform local instability analysis by solving the kinetic dispersion relation of whistler waves (see instability analysis in Methods), and find positive growth rate near the observed wave frequency and wave vector (Fig. 3h), indicating that the chorus waves were locally generated by the anisotropic electrons.

**Electron dynamics inside the chorus waves**

We now investigate electron dynamics in association with the chorus waves. As mentioned earlier, the waves are related to locally enhanced electron perpendicular temprature anisotropy which can seed the wave generation. According to cyclotron resonant condition in the non-relativistic regime, $\omega - k_\parallel V_{e,\parallel} = \omega_{ce}$, we estimate the resonant electron speed $V_{c,\parallel}$, yielding $V_{c,\parallel} = 3.7 \times 10^4$ km s$^{-1}$, or equivalently 4027 eV, relatively higher than local electron temperature of ~1800 eV. Aided by the unprecedentedly high-cadence measurements of MMS, we are able to study experimentally, for the first time, three-dimensional (3D) electron distribution functions inside the chorus waves, as displayed in Figure 4.

At the energy range close to the resonant energy, electron flux mainly occurs at pitch angles close to 90°, forming the so-called pancake distribution (Fig. 4D). While at relatively lower (< 3 keV) and higher energies (> 8 keV), electron fluxes do not show clear concentration near 90° (Fig. 4c and 4e). The presence of pancake distribution near the resonant energy further supports the chorus generation via cyclotron resonance.

Electron 3D velocity distribution functions (VDFs) inside the chorus waves are further examined. We focus on one rising-tone element (from ~00:34.7 to ~00:34.9 UT) which hosts strong wave intensity. Shown in Figs. 4f1-4i3 are two-dimensional, orthogonal cuts of the electron VDFs in $V_{e,\perp 1}-V_{e,\parallel}$, $V_{e,\perp 2}-V_{e,\parallel}$, and $V_{e,\perp 2}-V_{e,\perp 1}$ planes, where ∥ and ⊥ are directions parallel and perpendicular to ambient magnetic field, respectively. Electron resonant velocity $V_{c,\parallel}$ is also displayed inside the VDFs. Around $V_{c,\parallel}$, electron phase space density (PSD) shows significant variations and form distinct gradients therein, indicative of strong wave-particle interactions. Interestingly, local depletion of electron PSDs is also seen near $V_{c,\parallel}$, suggesting possible development of electron holes. In the electrons' cyclotron plane ($V_{e,\perp 1}-V_{e,\perp 2}$), the thermal core ($V<2\times10^4$ km s$^{-1}$) is basically isotropic, while weak nongyrotropy is seen at much higher energies ($V>4\times10^4$ km s$^{-1}$), in line with the significant variations of electron PSDs seen near $V_{c,\parallel}$.

Recent theories and simulations have suggested a crucial role played by electron holes in the wave phase space during chorus generation[25,26,33]. But they have not been verified by observations so far. Now we examine their existence using the MMS high-cadence data, focusing on the element (from ~00:34.7 to ~00:34.9 UT) studied above. Figure 5 displays the electron distributions in the wave phase space $(\zeta, V_\parallel)$, where $\zeta$ is the relative phase angle between the wave magnetic field and the electron perpendicular velocity with respect to ambient static magnetic field. At the resonant velocity (marked by the purple lines), we observe dramatic variations of electron distributions, indicative of wave-electron interactions. In addition, local depletions of electron PSD (marked by black dashed ellipses) are observed near the resonant velocity. These PSD depletions can be the electron holes predicted by recent theories and simulations.

**Energy transfer inside the wave**

The above analysis suggests that the chorus waves were growing, at the expense of electrons' energy. Now we directly examine the energy exchange between the chorus and the electrons in the plasma rest frame, based on Poynting's law (see energy transfer inside waves in Methods). The energy exchange inside the waves is quantified by $\delta E \cdot \delta j$, where $\delta E$ is the wave electric field and $\delta j$ is turbulent electron currents developed in associated with the waves. Here $\delta j$ is calculated using two different methods: Curlometer based on Ampere's law using MMS tetrahedron *(38)* (Fig. 6c), and direct estimate using electron moments from MMS1 ($\delta j_e$, Fig. 6d), which yield similar results. During the wave interval, $\delta E \cdot \delta j$ is basically negative (Figs. 6e), indicating ongoing energy transfer from the electrons to the waves. This direct measurement further confirms local generation of chorus waves by electron motions.

Another notable feature is that the energy transfer rate is correlated with the wave intensity. Shown in Fig. 6f is the correlation between the chorus wave power and the magnitude of local energy transfer rate. The wave power is positively correlated with the energy transfer rates, with Pearson correlation coefficient approaching 0.7, suggesting a strong correlation between the two. This indicates that stronger energy transfer yields more intense chorus, consistent with the ongoing excitation of the waves.

**Discussion and summary**

Chorus waves have so far been studied mainly in the near-Earth region ($L$-shells $\leq 9$), and have been widely believed to be governed by the magnetic dipolar field. Hence, the presence of repetitive, rising-tone chorus waves in the mid tail ($L$~25) where magnetic dipolar field effect is absent, provides new, important implications for understanding their generation and propagation in space. Interestingly, the observed rising-tone chorus waves have durations close to 0.1s, and chirping rates close to 250 Hz/s, close to those observed in the radiation belt. Considering that magnetic field and plasma populations in radiation belts and midtail neutral sheet are quite different, such similarities of chorus properties indicate that their generation is not uniquely determined by local environment therein.

Previous studies have suggested the crucial role played by relativistic effects in the chorus dynamics[25,26,28]. For the chorus waves observed in the mid tail, however, such relativistic effects are negligible, since the resonance velocity is much smaller than the speed of light. Hence relativistic effects are not critical for chorus generation. In addition, strong temperature anisotropy, with typical values close to 5, has been often assumed in previous studies. In the present case, the observed temperature anisotropy is much smaller, close to 1.2. These two differences suggested that previous knowledge of chorus generation needs to be modified.

The first measurement of electron 3D distributions inside the waves provides some evidence for previous theoretical predictions. In addition to the dramatic variations seen near the resonance velocity, the presence of electron holes in the wave phase space may be inspiring. These electron holes associated with chorus waves have been well studied by simulations[25,28,31], but it is the first time that they have been seen in observations. In addition, the energy exchange between the waves and local electrons, as well as the correlation between the wave amplitudes and energy conversion rates, support ongoing generation and growth of chorus waves via electron cyclotron resonance. The correlation between the wave amplitudes and energy conversion rates, however, does not support previous theoretical predictions, suggesting that the chirping rate is correlated with the wave amplitude (since chirping rate does not vary during the wave interval in the current observations)[29].

In short summary, this study presents the first observation of repetitive, rising-tone chorus emissions in the terrestrial mid tail, suggesting that chorus waves may be ubiquitous in the planetary magnetospheres, and provides direct measurement of wave-particle energy transfer inside the waves, showing evidence of strong wave-particle interactions. The results not only establish new pathways for studying chorus waves, i.e., using state-of-the-art measurements provided by MMS to diagnose electron dynamics inside chorus waves, but also provide new, important implications for understanding nonlinear wave-particle interactions in space and astrophysical plasmas.

## Methods

**Wave dispersion relation.**

The wave dispersion relation is resolved based on Ampere's law[36], which suggests that the relation between the wave vector and the wave magnetic field is:

$$\mu_0 \mathbf{J} = i\mathbf{k} \times \mathbf{B} \qquad (1)$$

where $\mathbf{J}$, $\mathbf{k}$, $\mathbf{B}$, and $\mu_0$ are wave current density, wave vector, wave magnetic field, and magnetic permeability in empty space, respectively. Based on this equation, we can directly calculate the magnitude and direction of $\mathbf{k}$ from single-spacecraft or multi-spacecraft measurements of current and magnetic field, without involving electric field data and assuming any dispersion relation for the observed waves. Here we have used four-point, high-cadence measurements provided by MMS, which are more applicable in resolving space-time ambiguity of the waves.

**Instability analysis.**

To investigate the chorus generation, we solve the kinetic dispersion relations of electromagnetic waves by using two different methods: one is widely-used WHAMP (Waves in Homogeneous Anisotropic Multicomponent Magnetized Plasma), and the other is recently-developed BO[39]. The main difference of these two methods is that WHAMP requires an initial guess for the wave dispersion relation, while BO can give all possible solutions without any initial guesses. The local plasma parameters in association with the observed wave emissions are used as inputs: for one rising-tone element studied: $|B|\sim 5\ nT, n_e \sim 0.1\ cm^{-3}, T_e \sim 5000\ eV, T_{e,\perp}/T_{e,\parallel} \sim 1.2$ ). The two methods yield similar positive growth rates, suggesting that the local conditions are indeed unstable to parallel-propagating whistle-mode waves.

**Energy exchange inside the waves.**

Using multipoint, high-cadence measurements provided by MMS mission, we investigate energy exchange inside the chorus waves, by examining the Poynting's law:

$$\frac{\partial E_{EM}}{\partial t} = -\mathbf{E} \cdot \mathbf{J} - \nabla \cdot \mathbf{S} \qquad (2)$$

$$\mathbf{S} = \mathbf{E} \times \mathbf{B}/\mu_0 \qquad (3)$$

where $E_{EM}$, $\mathbf{E}$, $\mathbf{J}$, $\mathbf{S}$, $\mathbf{B}$, $\mu_0$ represent electromagnetic field energy density, electric field, current density, Poynting flux density, magnetic field, and magnetic permeability in empty space, respectively. We decompose electric fields and currents into mean and fluctuating values with bandpass filtering (10-30 Hz in the present case):

$$\mathbf{E} \cdot \mathbf{J} = (\langle \mathbf{E} \rangle + \delta \mathbf{E}) \cdot (\langle \mathbf{J} \rangle + \delta \mathbf{J}) = \langle \mathbf{E} \rangle \cdot \langle \mathbf{J} \rangle + \delta \mathbf{E} \cdot \delta \mathbf{J} \quad (4)$$

where $\langle \mathbf{E} \rangle$, $\delta \mathbf{E}$, $\langle \mathbf{J} \rangle$, $\delta \mathbf{J}$ denote mean electric field, fluctuating electric field, mean current density, and fluctuating current density, respectively. Here the first term on the right side of Equation (4) quantifies laminar energy transfer, which is related to large-scale (at and above ion inertial scale) motion of particles and hence not studied here. The second term evaluates turbulent energy transfer (note that $\langle \mathbf{E} \rangle \cdot \delta \mathbf{J}$ and $\delta \mathbf{E} \cdot \langle \mathbf{J} \rangle$ terms turn into zeroes when being integrated in spatial domain and thus ignored) inside the chorus waves and is the focus of the present study.

## Data availability

All the data used in the present study is publicly available from the MMS Science Data Center (https://lasp.colorado.edu/mms/sdc/public/).

## Code availability

All of the data plots in this study are generated with the IRFU-Matlab software applied to the publicly available MMS database. The IRFU-Matlab software is available by downloading from https://github.com/irfu/irfu-matlab.

## Acknowledgements

We greatly appreciate the MMS Science Data Center for providing the data and IRFU-MATLAB for providing the analysis codes for this study. This research was supported by NSFC grants 42104164 and 41874188, and young talent supporting project of CAST (China Association for Science and Technology).

## Author contributions

C.M.L conducted the data processing, contributed to the data analysis, and prepared the manuscript. J.B.C oversaw the research project and gave suggestions. B.N.Z contributed to the data processing. J.L led the MMS mission. C.J.P, C.T.R. P.L contributed to the instrument development, operation, and data processing. C.J.P, C.T.R, Y.Y.L, and X.N.X gave suggestions on the manuscript writing. All authors reviewed the manuscript.

## Additional information

**Supplementary Information** is linked to the online version of the paper at www.nature.com.

**Competing interests**: The authors declare no competing interests.

**Author Information:** Correspondence and requests for materials should be addressed to jbcao@buaa.edu.cn

**Figure Legends**

**Fig. 1. Schematics showing the occurrence of whistler-mode chorus waves in geospace**. Chorus waves have hitherto been observed only in the near-Earth region, where magnetic field is dominated by dipolar field. The dipolar field introduces strong inhomogeneities of global magnetic field which has been believed to be closely related to the generation and propagation of the chorus waves. In the current study, MMS observed for the first-time chorus wave in the mid tail neutral sheet, where magnetic field is dominated by long, stretched field rather than dipolar field.

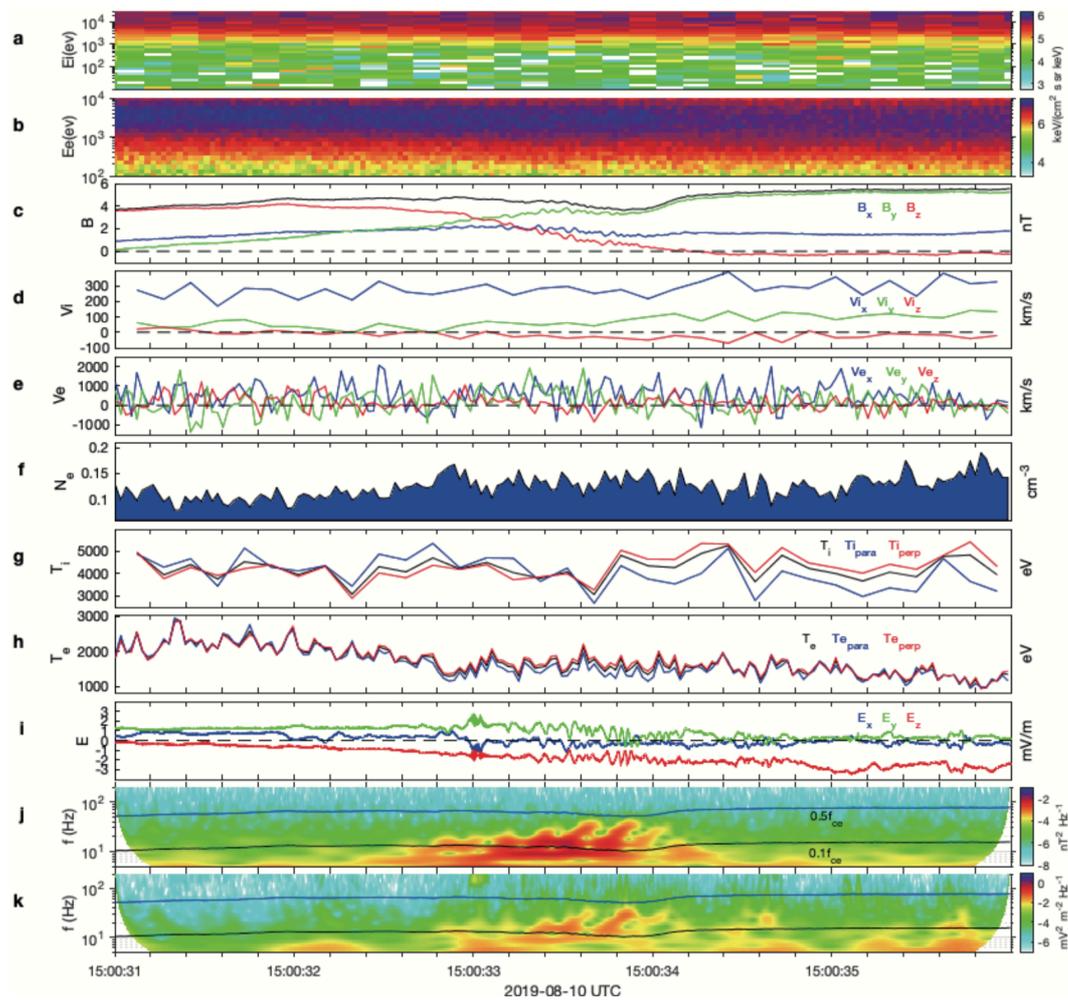

**Fig. 2. MMS observations of chorus waves in the midtail neutral sheet**. **a**, ion energy spectrum of differential energy fluxes. **b**, electron energy spectrum of differential energy fluxes. **c**, magnetic field. **d**, ion velocity. **e**, electron velocity. **f**, electron number density. **g**, ion temperature. **h**, electron temperature. **i**, electric field. **j**, power spectral density of magnetic field. **k**, power spectral density of electric field. The blue and black lines denote $0.5f_{ce}$ and $0.1f_{ce}$, respectively.

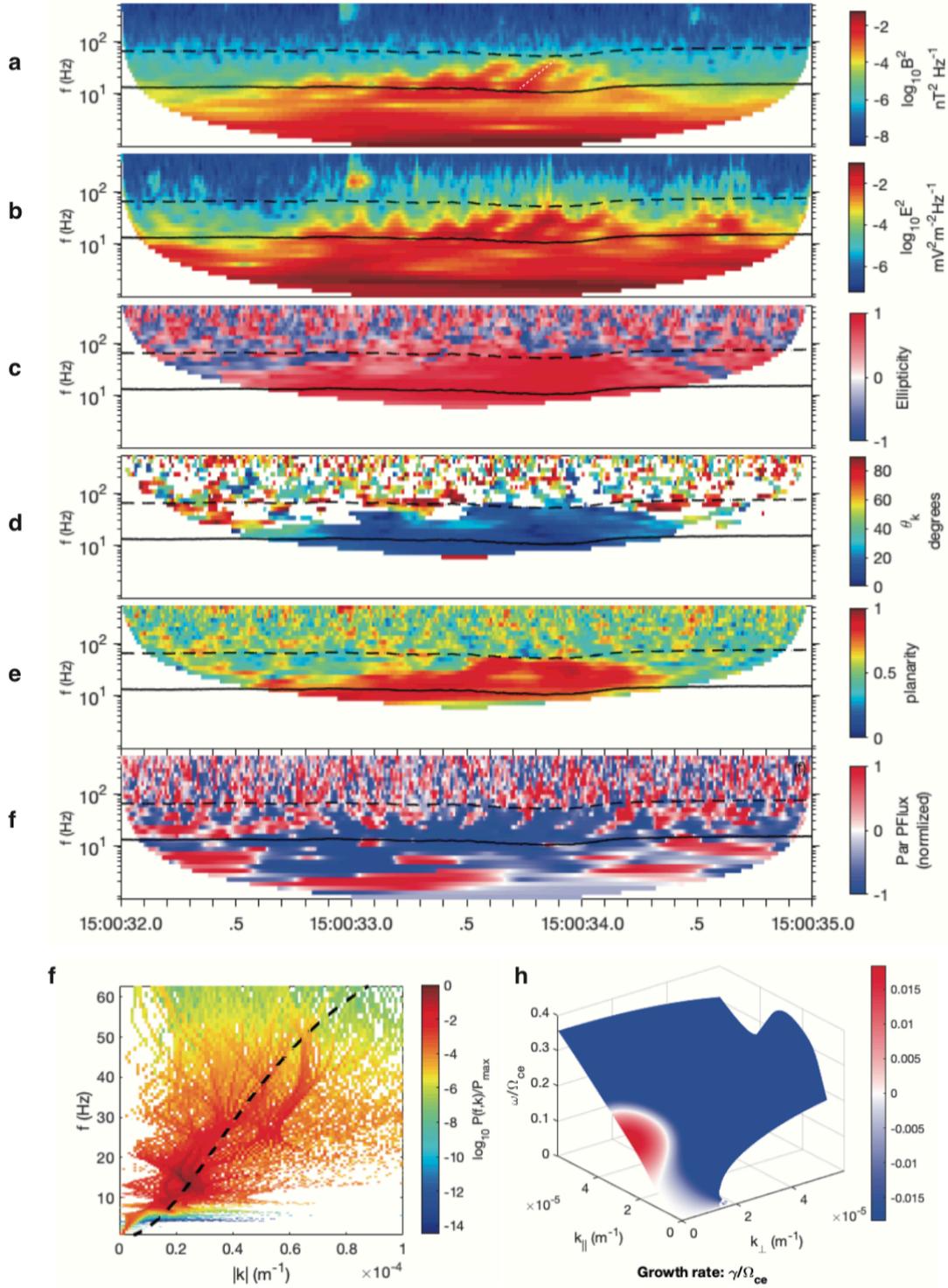

**Fig. 3. Wave polarization and dispersion relation**. **a**, power spectral density for magnetic field. **b**, power spectral density for electric field. **c**, ellipticity. **d**, wave propagation angle with respective to the magnetic field. **e**, planarity. **f**, normalized Poynting flux parallel to magnetic field. **g**, wave dispersion relation resolved via Ampere's law and its comparision with the thoretical dispersion relation for whistler

waves (black dashed line), and **h**, predicted dispersion surface of waves from WHAMP. The black-dashed and black-solid lines denote $0.5f_{ce}$ and $0.1f_{ce}$, respectively. The white dashed line in a denotes the frequency chirping of the waves.

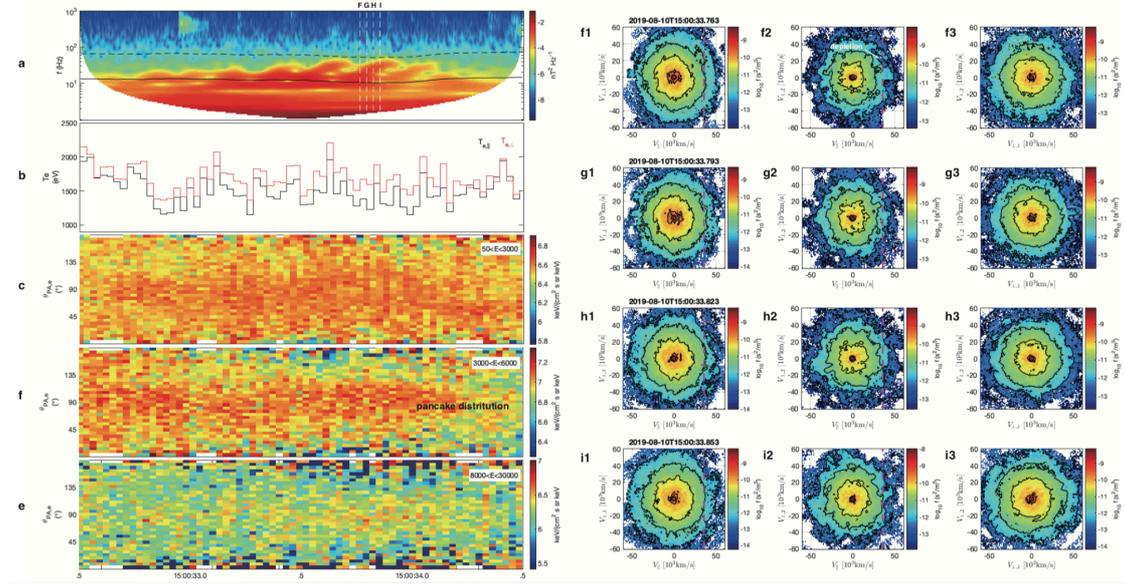

**Fig. 4. Electron distributions inside the chorus waves**. **a**, power spectral density of magnetic field. **b**, electron temperature. **c-e**, pitch angle distribution of electrons at low (50-3000 eV), mid (3-5 keV), and high energies (8-30 keV). **f1-i3**, 2D, orthogonal cuts of the electron 3D velocity distribution functions in the $V_{e,\perp 1}-V_{e,\parallel}$, $V_{e,\perp 2}-V_{e,\parallel}$ and $V_{e,\perp 1}-V_{e,\perp 2}$ planes, where $\parallel$ and $\perp$ are directions parallel and perpendicular to local magnetic field at times marked by white lines in A. The resonant speed is maked by purple dashed lines in panels f-i.

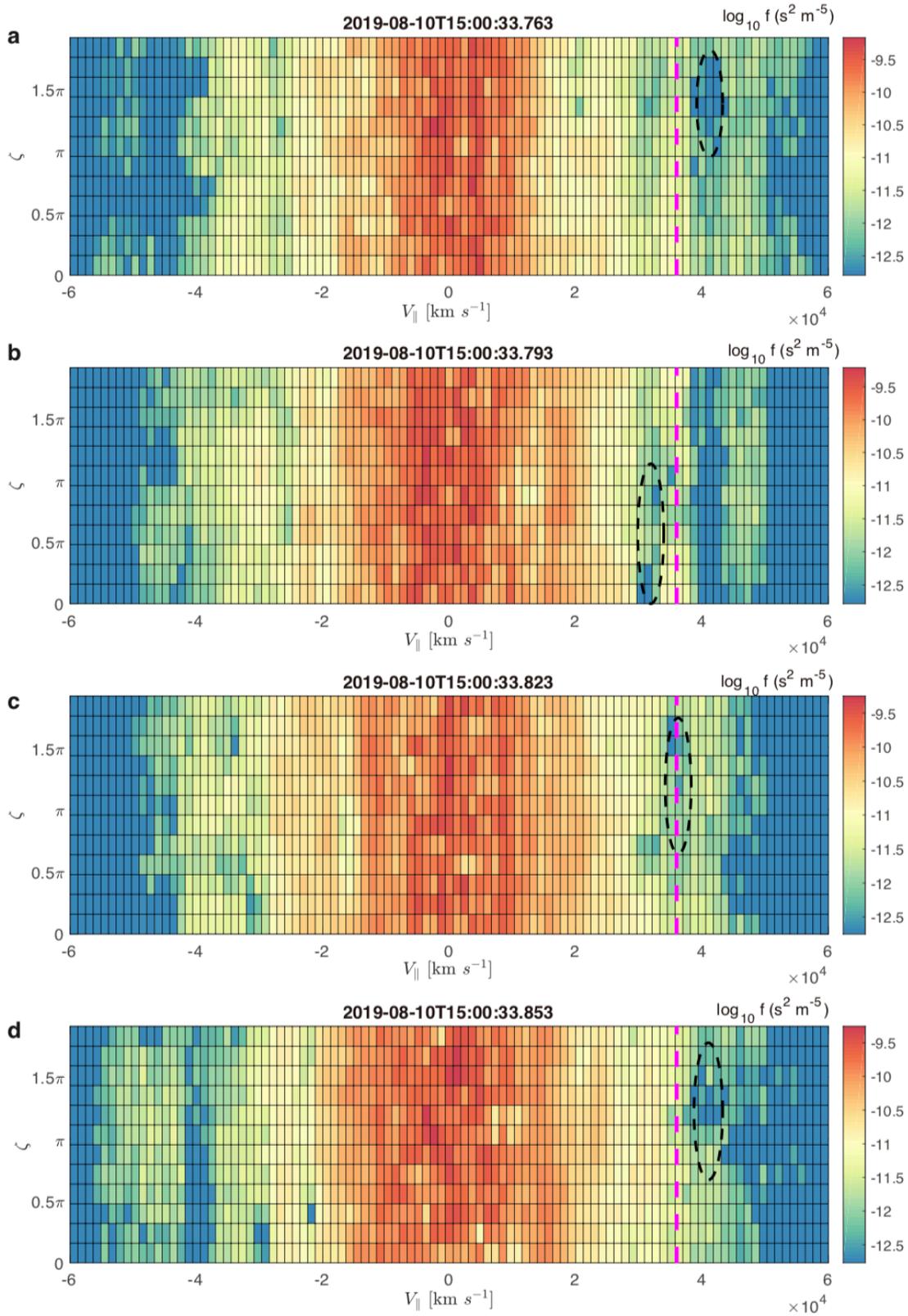

**Fig. 5. Electron distributions in the phase space of the chorus waves**. **a-d**, electron phase space density in the ($\zeta$, $V_\parallel$) phase space, where $\zeta$ is the relative phase angle between the wave magnetic field and the electron perpendicular velocity with respect

to ambient static magnetic field. The resonant speed is marked by purple dashed lines, the local depletion of electron distribution is marked by black, dashed ellipses.

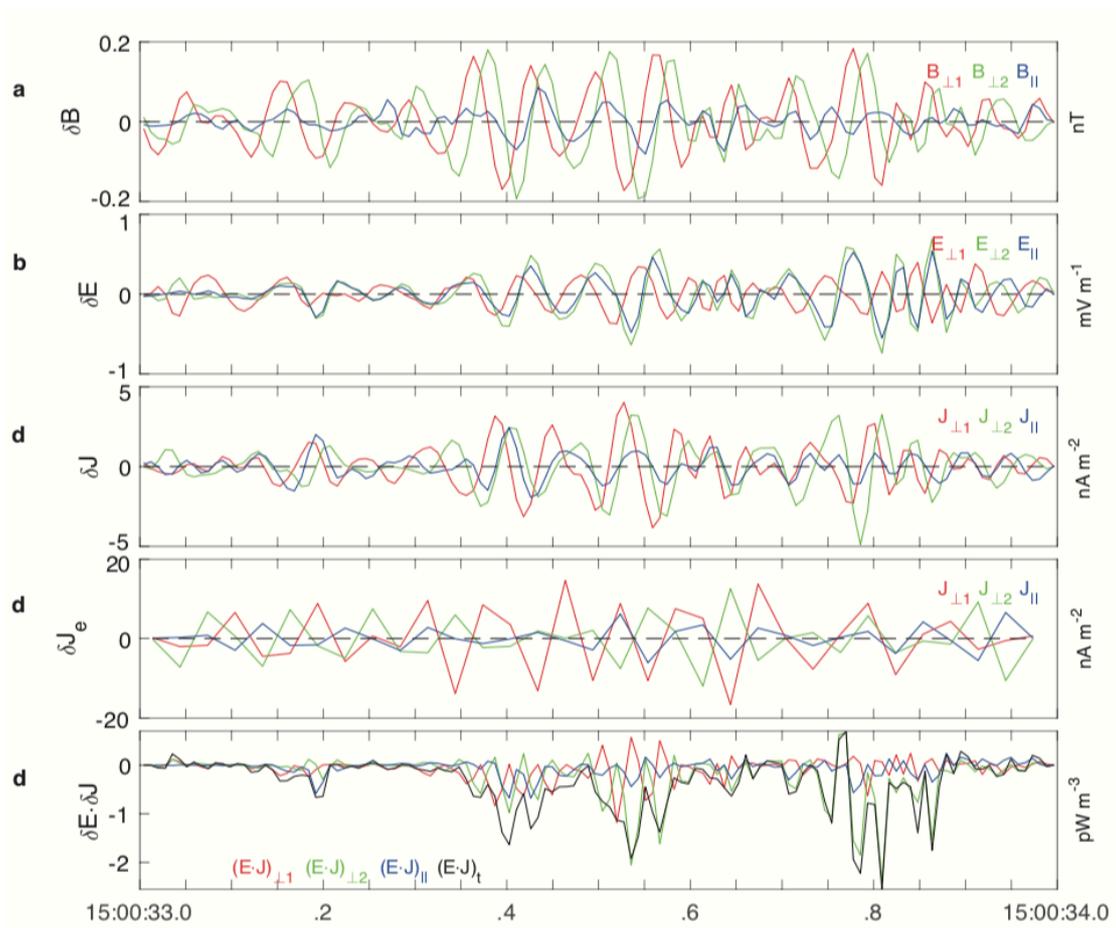

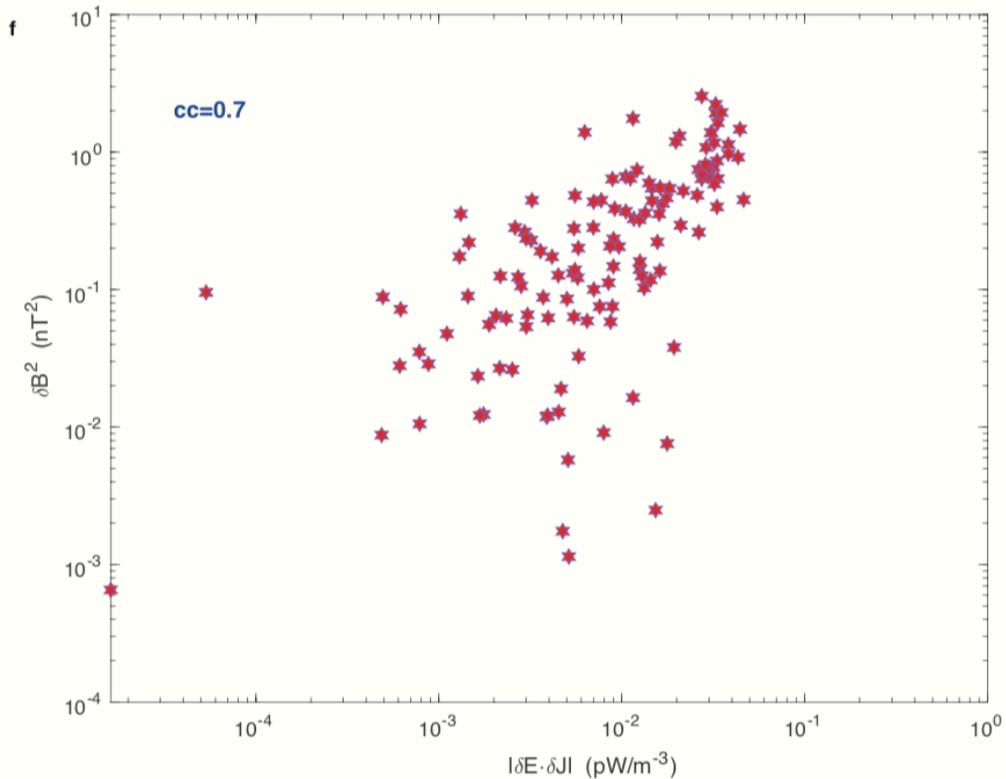

**Fig. 6. Turbulent energy exchange inside the chorus waves**. **a**, wave magnetic field. **b**, wave electric field. **c**, wave current calculated from Ampere's law. **d**, wave current calculated from electron moments. **e**, turbulent energy exchange inside the waves. **f**, correlation between wave amplitude and energy exchange rate. The wave amplitude is represented by $\delta B^2$ ($\delta B$ is the wave magnetic field), and the energy exchange rate is represented by $|\delta E \cdot \delta J|$, where $\delta E$ and $\delta J$ is wave electric field and current, respectively. The Pearson correlation coefficient is calculated, which approaches 0.7, indicating a nice correlation between the two parameters.